\documentclass [preprint] {revtex4}

\usepackage[dvips]{graphicx}
\usepackage{amssymb}
 \usepackage[latin1]{inputenc}

\def\beq {\begin{equation}}
\def\eeq {\end{equation}}
\def\bea {\begin{eqnarray}}
\def\eea {\end{eqnarray}}

\def\nn {\nonumber}

\def\lp {\left( }
\def\rp {\right) }
\def\lb {\left[ }
\def\rb {\right] }
\def\lc {\left\{ }
\def\rc {\right\} }
\def\ra {\;\rangle }
\def\la {\langle\; }

\def\rar {\rightarrow}

\def\cb {\bar{c}}

\def\sb {\bar{s}}

\def\sb {\bar{s}}

\def\Kb {\bar{K}}

\def\Ob {\bar{\Omega}}

\def\Tb {\bar{T}}

\def\ct {\tilde{c}}

\def\dkpp {$D^+ \! \rar \! K^- \p^+ \p^+$ }

\def\sp {\!+\!}
\def\sm {\!-\!}
\def\cd {\!\cdot\!}

\def\cK {{\cal{K}}}

\def\cO {{\cal{O}}}

\def\a {\alpha }
\def\b {\beta}
\def\d {\delta}
\def\D {\Delta}
\def\e {\epsilon}

\def\g {\gamma}
\def\l {\lambda }

\def\m {\mu}
\def\n {\nu}

\def\p {\pi}
\def\P {\Pi}

\def\th {\theta}
\def\T {\Theta}

\def\bq {\mbox{\boldmath $q$}}

\begin{document}

\title{$D^+ \rar K^- \p^+ \p^+$ - the weak vector current}

\author{ P. C. Magalh\~{a}es}
\email[]{patricia@if.usp.br}
\author{M. R. Robilotta}
\affiliation{ Instituto de F\'{\i}sica, Universidade de S\~{a}o Paulo,  
S\~{a}o Paulo, SP, Brazil}
 
\date{\today }

\begin{abstract}
Studies of  D and B mesons decays into hadrons
have been used to test the standard model in the last fifteen years.
A heavy meson decay involves the combined effects of a primary weak 
vertex and subsequent hadronic final state interactions,
which determine the shapes of Dalitz plots. 
The fact that final products involve light mesons indicates
that the QCD vacuum is an active part of the problem.
This makes the description of these processes rather involved and, 
in spite of its importance, 
phenomenological analyses tend to rely on crude models. 
Our group produced, some time ago,  a schematic calculation of
the decay $D^+ \rar K^- \p^+ \p^+$, which provided a reasonable 
description of data.
Its main assumption was the dominance of the weak vector-current,
which yields a non-factorizable interaction. 
Here  we refine that calculation by including the
correct momentum dependence of the weak vertex and extending
the energy ranges of $\p\p$ and $K\p$ subamplitudes
present into the problem.
These new features make the present treatment more realistic and
bring theory closer to data.

\end{abstract}

\pacs{...}

\maketitle

\section{motivation}

Non-perturbative QCD calculations are difficult and can only be 
performed in approximate frameworks.
The grouping of quarks into two sets, 
according to their masses, provides
a convenient point of departure for approximations.
Quarks $u$, $d$, and $s$ can be considered as light 
and quarks $c$, $b$, and $t$, as heavy, even though the $s$-quark is not 
too light and the $c$-quark is not too heavy.
This approach is useful because light quark condensates are 
active close to the ground state of QCD and give rise to
highly collective interactions.

Pions and kaons are the most prominent light  quark systems,
but data available for 
elastic $K\p$ scattering are scarce and decades old.
They were obtained from the LASS spectrometer at SLAC\cite{LASS,Estab},
in the range $0.825 < \sqrt{s} < 1.960$ GeV,
by isolating one-pion exchanges in
the reaction $KN \to \pi KN$.
In the last ten years, information about $K\p$ interactions 
was also produced by hadronic decays of $D$ mesons.
In particular, data from the E791 and FOCUS 
collaborations\cite{E791kappa, FOCUS}
for the reaction $D^+ \to K^-\pi^+\pi^+$ allowed the
$S$-wave $K\p$ sub-amplitude to be extracted
continuously from threshold up to the high energy border of the 
Dalitz plot.
Hope was then raised that these data could improve
the description of elastic $K\p$ scattering.
However, decay data differ significantly 
from those given by the LASS experiment and 
this discrepancy motivates our interest in this problem.

The description of the decay $D^+ \to K^-\pi^+\pi^+$
must include both the weak $c \rar s$ vertex
and hadronic final state interactions (FSIs), 
which correspond to strong processes occurring between 
primary decay and detection.
The study of weak vertices departs from the topological structures 
given by Chao\cite{Chao},
which implement CKM quark mixing 
for processes involving a single $W\,$.
As primary decays occur in the presence of light quark condensates,
the direct incorporation of Chao's scheme into calculations is 
not trivial and
one is forced into hadronic descriptions. 
These include both the use of form factors in weak vertices,
as in the work of Bauer, Stich and Wirbel\cite{BSW},
and the treatment of relativistic final state interactions.
High-energy few-body calculations begin to be available 
now\cite{Azimov,lc09,Zhou}
and several works have already employed field theory to FSIs
in heavy meson decays\cite{Ca,Bo,Me,Jap,DeD,DiogoRafael,BR, satoshi}.

In this work, the decay $D^+ \to K^-\pi^+\pi^+$ is treated 
by means of chiral effective lagrangians, 
supplemented by phenomenological form factors.
This framework is motivated by the smallness of the
$u\,$, $d\,$, and $s$ masses, when compared with the 
QCD scale $\Lambda \sim 1\,$GeV.
The light sector of the theory is therefore not far from 
the massless limit,
which is symmetric under the chiral $SU(3)\times SU(3)$ flavour group.
In this approach, light condensates arise naturally
and pseudoscalar mesons are described as Goldstone bosons.
Quark masses are incorporated perturbatively
into effective lagrangians\cite{Wchi,GL},
whereas weak interactions are treated as external sources.
Chiral perturbation theory was originally designed to 
describe low-energy interactions, where it yields 
the most reliable representation of QCD available at present.
Its scope was later enlarged, with the inclusion of resonances
as chiral corrections\cite{EGPR}, and the unitary ressummation
of diagrams\cite{OO}.  
Suitable coupling schemes also allow the incorporation
of heavy mesons\cite{HM}.

A similar theoretical framework has already been employed 
by our group\cite{BR}, in an exploratory study of FSIs 
in $D^+ \rar K^- \p^+ \p^+$.
With the purpose of taming an involved calculation,
in that work we made a number of simplifying assumptions.
Among them, the weak vertices were taken to be constants,
isospin $3/2$ and $P$ waves were not included in 
intermediate $K\p$ amplitudes,
and couplings to either vector mesons or 
inelastic channels were neglected.
In spite of these limitations, that work allowed the identification
of leading dynamical mechanisms and gave rise to results
which are reasonable for the modulus and good for 
the phase of the $S$-wave $K\p$ sub-amplitude\cite{E791kappa,FOCUS}.
In this work, we focus on the vector weak amplitude and improve  the description of the weak vertex,
by including both the correct momentum dependence
and better phenomenology for an intermediate $\p\p$
subamplitude, and the description of a $K\p$
subamplitude at higher energies.
These new features tend to reduce the gap between theory 
and experiment.

\section{dynamics}

\begin{figure}[h]
\hspace*{-10mm}
\includegraphics[width=.9\columnwidth,angle=0]{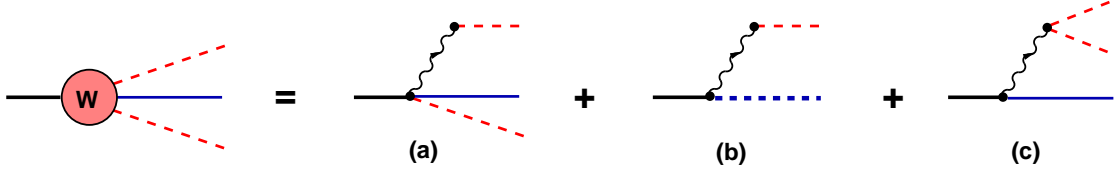}
\caption{Topologies for the weak vertex: the dotted line is a scalar 
resonance and the wavy line is the $W^+$, which is contracted to a point 
in calculations.}
\label{FAmpW}
\end{figure}

We denote by $[K^-\p^+]_S$, the $S$-wave $K^- \p^+$ sub-amplitude 
in the decay 
$D^+ \rar K^- \p^+ \p^+$, which has been extracted 
by the E791\cite{E791kappa} 
and FOCUS\cite{FOCUS} collaborations.
The decay begins with the primary quark transition $c\rar s \;W^+$,
which is subsequently dressed into hadrons,
owing to the surrounding light quark condensate.
In the absence of form factors, this structure gives rise to the 
colour allowed process shown in Fig.\ref{FAmpW}, 
where $a$ and $b$ involve an axial current and $c$
contains a vector  current.
As one of the pions in diagram $c$ is neutral,
it does not contribute at tree level.

\begin{figure}[h]
\includegraphics[width=.9\columnwidth,angle=0]{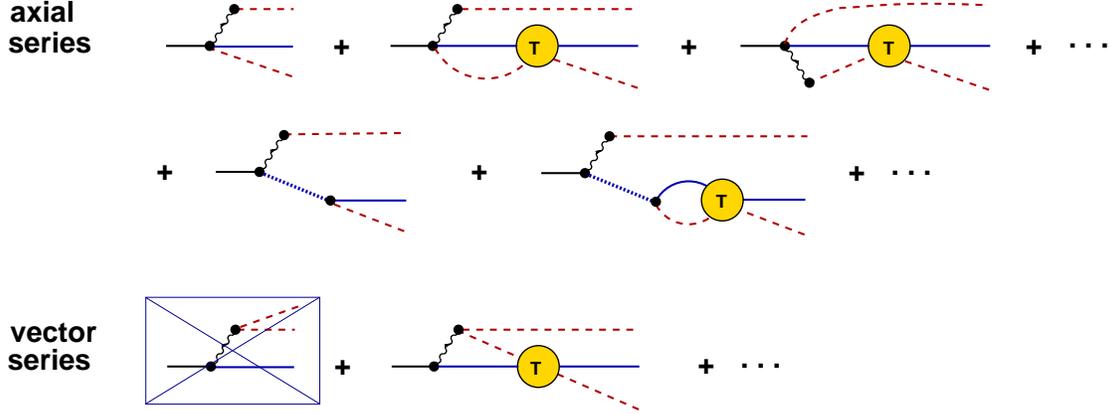}
\caption{Final state interactions starting from the axial weak vertex
(axial series ) and from the vector weak vertex (vector series);
in the former, the pion plugged to the $W^+$ is
always positive, whereas the $\Kb$ inside the loop can be either 
positive or neutral;
in the latter, the tree diagram does not contribute, since one of the
pions plugged to the $W^+$ is neutral.}
\label{FSI}
\end{figure}

Inclusion of final state interactions, due to successive 
elastic $K\p$ scatterings\cite{BR}, yields 
three families of diagrams, as in Figs.\ref{FSI}.
It is worth noting that these series {\em do not} represent
a loop expansion, because loops are also present within
the $K\p$ amplitude.
The $W^+$ is shown explicitly, just to indicate the 
various topologies, and becomes point-like in calculations.
A family of FSIs endows the forward propagating resonance in Fig.1b with a dynamical width\cite{Diogo}.
Processes involving resonances have already been 
considered in Refs.\cite{DeD,Jap,satoshi},
whereas quasi two-body axial FSIs were discussed Ref.\cite{DiogoRafael}.
An important lesson drawn from our previous study\cite{BR}
is that, for some yet unknown reason, the vector weak amplitude, 
represented by diagram $(c)$ of Fig.\ref{FAmpW}, 
seems to be favoured by data\cite{FOCUS}.
This amplitude receives no contribution at tree level, since the
$W^+$ emitted by the $c$-quark decays into a $\p^+\p^0$ pair.
Therefore, leading terms in this process necessarily
involve loops, which bring imaginary components into the amplitude.

\begin{figure}[h] 
\hspace*{-30mm}
\includegraphics[width=0.5\columnwidth,angle=0]{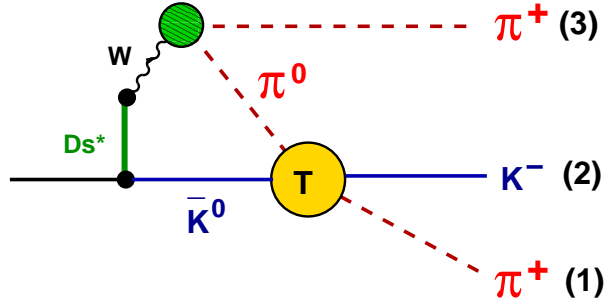}
\caption{Leading vector current contribution, dressed by 
form factors and $\p\p$ 
interactions (in the small green blob).}
\label{Fvec}
\end{figure}
%

The first non-vanishing contribution to the vector series is 
given in Fig.\ref{Fvec}.
As the $W$ is very heavy, one keeps just hadronic propagators, 
which render loop integrals finite.
Denoting by $A_0$ the amplitude for the process 
$D^+ \to K^0\pi^0\pi^+$ without FSIs and by $T_{K\p}$ that for  
$\p^0 \Kb^0 \to \p^+ K^-$,
the amplitude $A_1$ of Fig.\ref{Fvec} 
can be schematically written as
\bea
A_1= -i \int \frac{d^4 \ell}{(2\p)^4} \; 
T_{K\p}^S \,\D_\p \,\D_K \; A_0\;,
\label{2.1}
\eea
where $\ell$ is the loop variable and 
$\D_\p$ and $\D_K$ are pion and kaon propagators.

\begin{figure}[h] 
\includegraphics[width=0.7\columnwidth,angle=0]{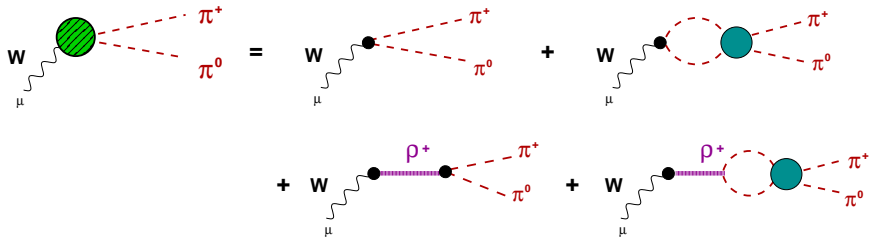}
\caption{Structure of the $W \rar \p \p$ form factor;
the blue blob is the elastic $\p\p$ amplitude.}
\label{Fro}
\end{figure}

The amplitude $A_0$ is described in App.\ref{basic}.
The $D \rar W \Kb$ vertex includes $D_s^*$ intermediate states,
associated with form factors 
parametrized in terms of nearest pole dominance \cite{weakFF} and could be a vector or a scalar.
The $W\rar \p\p$ form factor is shown in Fig.\ref{Fro}
and includes the $\rho$, with a dynamical width.
The bare resonance is treated employing
the formalism developed in Ref.\cite{EGPR} and 
its width is constructed using the $P$-wave 
elastic $\p\p$ amplitude.
The $W\rar \p\p$ form factor is time-like and 
its inclusion into the vector series of Fig.\ref{FSI} 
can, in principle, give rise to  
final state interactions depending  on both $\p\p$ and $K\p$
amplitudes.
With the purpose of keeping complications to a minimum,
we consider just $\p\p$ interactions which are contiguous
to the $W$ and occur to the left of the first $K\p$ amplitude.

The evaluation of Fig.\ref{Fvec} requires the $K\p$ amplitude in
the interval $0.401\,$GeV$^2 \leq s \leq 2.993\,$GeV$^2$.
As LASS data\cite{LASS} begins only at $s = 0.681\,$GeV$^2$,
one covers the low-energy region by means
of theoretical amplitudes, based on unitarized 
chiral symmetry\cite{EGPR}.
Our intermediate $S$-wave $K\p$ amplitude, 
denoted by $T_{K\p}^S$, is thoroughly discussed in App.\ref{kapi}.

Using results $(\ref{B.17})$ into  eq.$(\ref{2.1})$,
one finds 
\bea
A_1^S(m_{12}^2) &\!=\!& - \, i \, 
\lb G_F\, \cos^2\th_C  \, F_1^{DK}(0)\rb \; 
\lb \frac{\sqrt{2}}{3}\, T_{K\p}^S(m_{12}^2)\rb 
\int \frac{d^4 \ell}{(2\p)^4} \;
\frac{1}{D_\p \, D_K} \; \frac{m_\rho^2}{D_\rho} 
\nn\\[2mm]
&\! \times \!& \lc
[M_D^2 \sp 2M_\p^2 \sp M_K^2 \sm 2 m_{12}^2 \sm \ell^2 
\sp D_\p \sp D_K \,]\,\frac{m_V^2}{D_V}
\right.
\nn\\[4mm]
&\!+\!& \left.
D_\p \, (M_D^2 \sm M_K^2) \lb  \frac{1}{D_V} - \frac{1}{D_S} \rb \rc \;,
\label{2.2}
\eea
where $G_F$ is the Fermi constant, $\th_C$ is the Cabibbo angle,
$F_1^{DK}(0)$ is a coupling constant\cite{weakFF},
the factor $\sqrt{2}/3$ is associated with the 
transition $K^0 \p^0 \rar K^-\p^+$, 
whereas $D_\p = [(\ell\sm p_3)^2 \sm M_\p^2]$,
$D_K=[(\ell\sm P)^2 \sm M_K^2]$, $D_V = [\ell^2 \sm m_V^2]$,
$D_S=[\ell^2 \sm m_S^2]$, 
in which the subscripts $V$ and $S$ stand for
the $D_s^*(2112)$ and $D_{s0}^*(2317)$ states. Finally, $D_\rho$ is a complex function
defined by eqs.(\ref{B.15}) and (\ref{B.16}).
This structure yields 
\bea
A_1^S(m_{12}^2) &\!=\!& 
- i \, \a \,\frac{\sqrt{2}}{3}\, 
\lb \frac{T_{K\p}^S(m_{12}^2)}{16 \p^2} \rb   
\lc \; \b \,  
I_{\p K \rho V}^S - I_{\p K  V}^S + I_{\p \rho  V}^S
+ I_{K \rho V}^S - \g\, I_{K \rho V S}^S \rc \;,
\label{2.3}
\eea
with
\bea
\a &\!=\!& \lb G_F\, \cos^2\th_C \, F_1^{DK}(0)\rb \; 
m_\rho^2 \, m_V^2 \;,
\label{2.4}\\[2mm]
\b &\!=\!&  M_D^2 \sp 2M_\p^2 \sp M_K^2 \sm m_\rho^2 \sm 2 m_{12}^2 \;, 
\label{2.5}\\[2mm]
\g &\!=\!& - (M_D^2 \sm M_K^2)(m_V^2 \sm m_S^2)/m_V^2 \;,
\label{2.6}
\eea
and 
\bea
I_{abc}^S &\!=\!& \int \frac{d^4 \ell}{(2\p)^4} \;
\frac{16 \p^2 }{D_a \, D_b \, D_c } \;,
\;\;\;
I_{abcd}^S = \int \frac{d^4 \ell}{(2\p)^4} \;
\frac{16 \p^2 }{D_a \, D_b \, D_c \, D_d } \;.
\label{2.7}
\eea
The form of these integrals is discussed in App.\ref{int}.

\section{vector FSI series}

\begin{figure}[h] 
\hspace*{-20mm}
\includegraphics[width=0.8\columnwidth,angle=0]{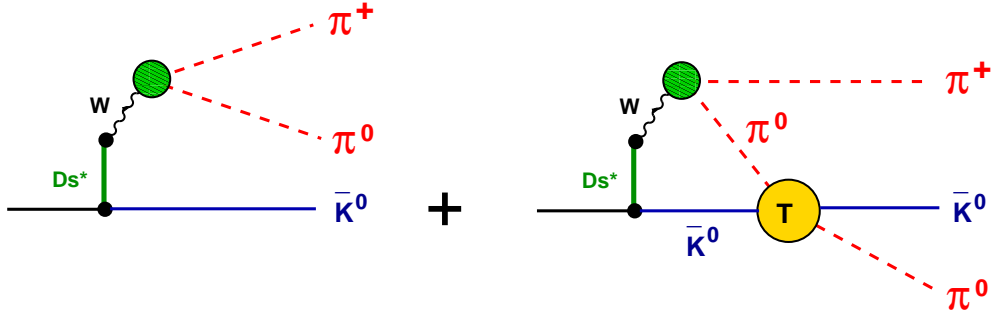}
\caption{Vector current diagrams contributing to the decay 
$D^+ \rar \Kb^0 \p^0 \p^+$.}
\label{FroG}
\end{figure}

\begin{figure}[ht]
\begin{center}
\includegraphics[width=0.6\columnwidth,angle=0]{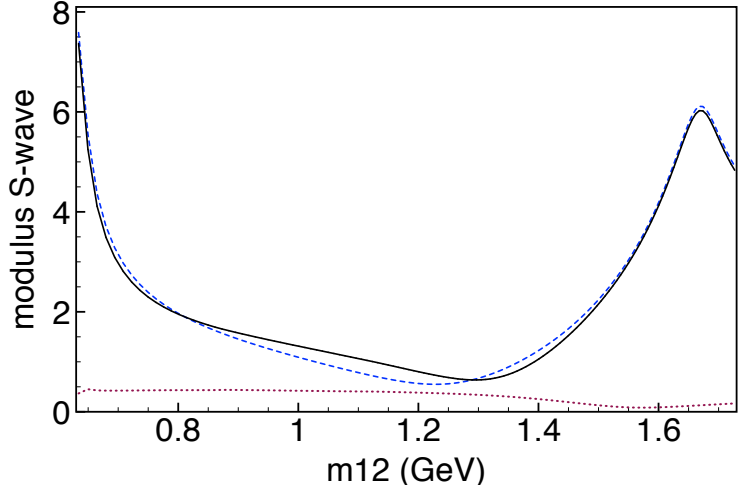}
\caption{Modulus of the $D^+ \rar \Kb^0 \p^0 \p^+$ amplitude
(full line) and partial contributions 
from eqs.(\ref{3.1}) (dashed line) and (\ref{3.3}) (dotted line).} 
\label{convergence}
\end{center}
\end{figure}

In the decay $D^+\rar K^-\p^+\p^+$, there is no tree
contribution to the vector FSI series, as in Fig.\ref{FSI}. 
However, before moving into this reaction, it is instructive
to assess the relative importance of allowed tree and one-loop 
contributions in the decay $D^+ \rar \Kb^0 \p^0 \p^+$,
indicated in Fig.\ref{FroG}.
The amplitude describing the left diagram is denoted by $A_0$
and given in eq.(\ref{B.17}).
Projecting out the $S$-wave, we find
\bea
A_0^S &\!=\!& -[G_F\, \cos^2\theta_C\,F_1(0)]
\sum_i 
\lc \frac{m_V^2 \,N_i}{m_V^2 \sm \th_i} \right.
\nn\\[2mm]
&\! \times \!& \left.
\lb \lp M_D^2  + 2M_\p^2 + M_K^2 - 2 m_{12}^2 - m_V^2 \rp \;\P_V 
\right.\right.
\nn\\[2mm]
&\! - \! & \left. \left.
\lp M_D^2  + 2M_\p^2 + M_K^2 - 2 m_{12}^2 - \th_i \rp \;\P_{\th_i}
\rb \rc \;,
\label{3.1}\\[2mm]
\P_{[V;\th_i]} &\!=\!& \frac{1}{2\,\b} \; 
\ln  \frac{[\,m_V^2; \,\th_i\,] \sm \a_{13}^2 \sm \b\,}
{[\,m_V^2; \,\th\,] \sm \a_{13}^2 \sp \b\,} \;,
\label{3.2}
\eea
where $\th_i$ are complex parameters given in table \ref{BT} (App.\ref{basic}). 
The first order amplitude is obtained by replacing the
isospin factor $\sqrt{2}/3$ with $-1/3$ in eq.(\ref{2.3}) and reads
\bea
A_1^S(m_{12}^2) &\!=\!& 
 i \, \a \,\frac{1}{3}\, 
\lb \frac{T_{K\p}^S(m_{12}^2)}{16 \p^2} \rb   
\lc \; \b \,  
I_{\p K \rho V}^S - I_{\p K  V}^S + I_{\p \rho  V}^S
+ I_{K \rho V}^S - \g\, I_{K \rho V S}^S \rc \;,
\label{3.3}
\eea
Results for the moduli of $A_0^S$ and $A_1^S$, 
displayed in Fig.\ref{convergence}, indicate a clear dominance 
of the former.
The main structural difference between both terms is 
the factor $\lc T_{K\p}^S/48 \p^2 \rc$ in the latter,
associated with a final state scattering.
Its scale can be understood by noting that
chiral symmetry predicts this amplitude to be
$T_{K\p}^S=2\,M_\p M_K/F^2 \sim 13$ at threshold whereas
LASS data \cite{LASS} indicate that it reaches a maximum of $T_{K\p}^S \sim 50$
around $m_{12} \sim 1.33\,$GeV.
Therefore the factor $\lc T_{K\p}^S/48 \p^2 \rc$ is always smaller 
than $1/10\,$ and pushes down the loop contribution.
This result can be taken as an indication that the
vector series, as given in fig.\ref{FSI}, converges rapidly.
The confirmation of this hint depends, of course, on the
explicit calculation of next terms in the series.

\section{results - $S$-wave}

\begin{figure}[ht!]
\includegraphics[width=0.7\columnwidth,angle=0]{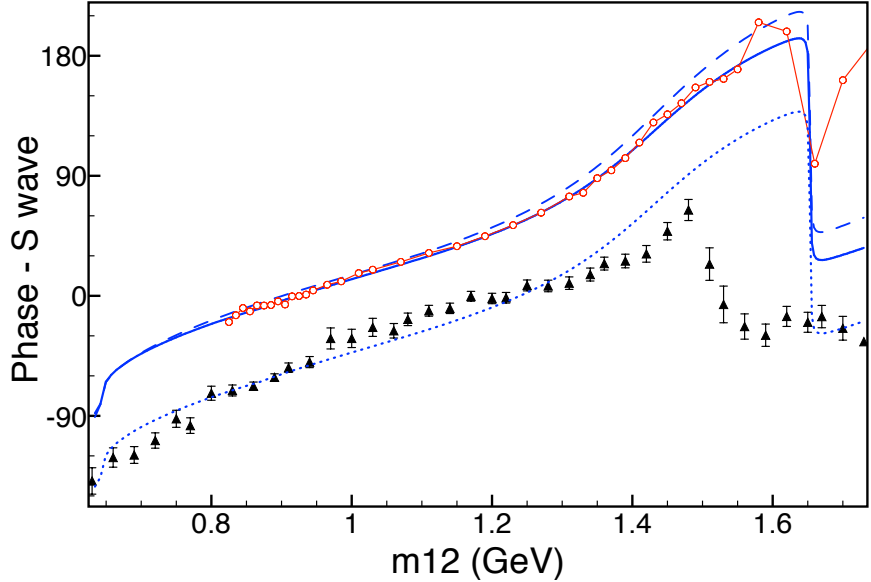}
\caption{Predictions for \dkpp phase (full blue curve),
based on the parametrized $\p\p$ and $K\p$ amplitudes given
in appendices \ref{basic} and \ref{kapi}, compared with 
FOCUS data\cite{FOCUS}; the blue dotted curve is the previous one shifted by $-55^0$; the dashed blue curve is based on one-$\rho$ pole approximation  for the $\p\p$ amplitude; 
in the  red symbol-continuous curve  the   {\em hybrid} model was used for the $K\p$ amplitude. 
}
\label{FPh}
\end{figure}

\begin{figure}[ht!]
\includegraphics[width=0.7\columnwidth,angle=0]{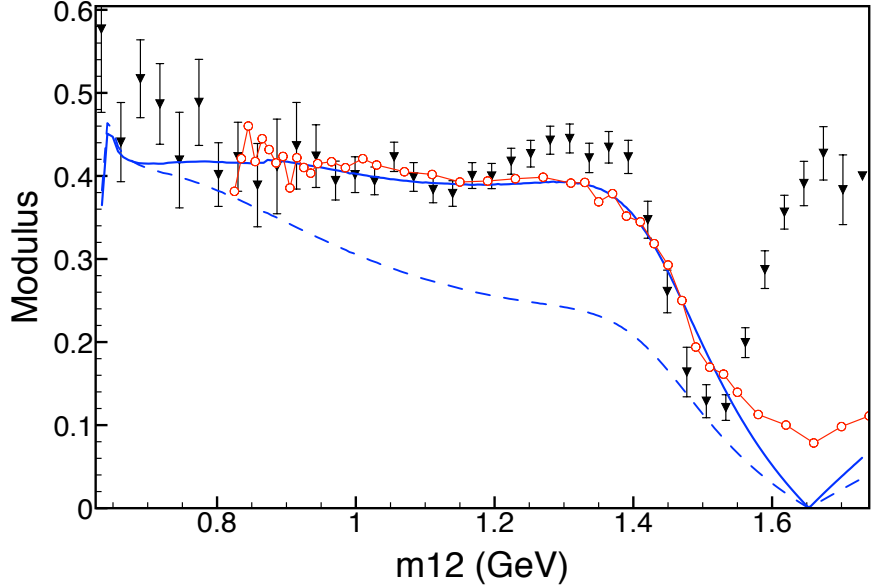}
\caption{Predictions for \dkpp modulus (full blue curve),
based on the parametrized $\p\p$ and $K\p$ amplitudes given
in appendices \ref{basic} and \ref{kapi}, compared with 
FOCUS data\cite{FOCUS}, using arbitrary normalization;
the dashed blue curve is based on one-$\rho$ pole approximation  for the $\p\p$ amplitude; 
in the  red symbol-continuous curve  the   {\em hybrid} model was used for the $K\p$ amplitude.}
\label{FMo}
\end{figure}

One of the purposes of this work is to understand 
the role played by the high energy components of 
intermediate $\p\p$ and $K\p$ subamplitudes in the description 
of data. Predictions from eq.(\ref{2.3}) for the phase and modulus 
of $[K^-\p^+]_S$, 
the $S$-wave $K^- \p^+$ sub-amplitude in $D^+ \rar K^- \p^+ \p^+$,
are given in Figs.\ref{FPh} and \ref{FMo}.

As far as the $\p\p$ subsystem is concerned, the data of 
Hyams et al.\cite{Hyams} are used in a parametrized form,
in the whole region of interest, as discussed in App.\ref{basic}.
For the sake of producing a contrast, we also show curves 
corresponding to the low-energy vector-meson-dominance
approximation,
in which the $P$-wave amplitude is described by 
just an intermediate $\rho$-meson, which  amounts to using just the first term in eq.(\ref{B.16}).
In the case of the $K\p$ amplitude, data are not available  for
energies below $0.825\,$GeV \cite{LASS} and two alternative extensions
are given in App.\ref{kapi}.
One of them is based on a two-resonance fit, which encompasses 
both low- and high-energy  sectors,
whereas in the other one LASS data\cite{LASS} is used directly,
when available, and extrapolated to the threshold region by 
means of a fit.
In the sequence we refer to these versions as
{\em fitted} and {\em hybrid}, respectively.
The main difference between them is that the former 
excludes points around $E \sim 1.7\,$GeV,
shown in Fig.\ref{FC1},
where two-body unitarity is violated.

Inspecting the figures, one learns that the improvement  
in $\p\p$ phenomenology is more important for the modulus, 
where it influences considerably the curve behaviour 
and increases significantly the range in energy where the
theoretical description proves to be reasonable. 
In the case of the phase, effects associated with $\p\p$ 
phenomenology are small and visible only above 
$m_{12}> 1.2\,$GeV. 
On the other hand, the use of either the fitted or hybrid
$K\p$ amplitudes produces equivalent results, except 
at the high energy end, where none of them is satisfactory.
This seems to indicate  missing structures,  that could be associated with other topologies in \dkpp decay.

As experimental results for the FOCUS phase\cite{FOCUS}
include an arbitrary constant, in Fig.\ref{FPh}
we also show our main result displaced by $-55^0$.
One notices an overall good agreement with data,
from threshold up to $m_{12}\sim 1.4\,$GeV.
As our  results were based on the vector series shown in Fig.\ref{FSI},
which does not contain a tree contribution,
 there are two sources of complex phases in 
this problem.
One of them is that associated with the $K \p$ amplitude,
whereas the other one is less usual and due to the loop
including the weak vertex.
Our results indicate that the latter is rather important 
over the whole energy range considered.
This shows the relevance of proper three body interactions,
which share the initial momentum with all final particles at
once.
\section{conclusions}

In this work we calculate the weak vector current contribution to the 
process \dkpp, employing  intermediate $\p\p$ and $K \p$ intermediate
sub-amplitudes valid within most of the Dalitz plot.
Together with the use of a proper $P-$wave weak vertex,
this extends a previous study made on the subject\cite{BR}.  
We still concentrate on $[K^-\p^+]_S$, 
the $S$-wave $K^- \p^+$ sub-amplitude,
and present predictions for both the phase and modulus,  given 
by the blue curves in Figs.\ref{FPh} and \ref{FMo}, are quite satisfactory from threshold to $1.4\,$GeV.
Results for the modulus, in particular,  improve considerably 
our previous findings, showing that intermediate $\p\p$
subamplitudes are important and need to be treated carefully.  
As far as the phase is concerned, the most prominent feature
is the fact that it has a large negative value at threshold.
In QCD, loops are the only source of complex amplitudes
and, in this problem, the energy available in the loop of 
Fig.\ref{Fvec} can be larger than both  $K \p$ and $K \rho$
thresholds.
This yields a rich complex structure for the loop 
containing the W, with  a phase $\Theta_L$  which  adds to the phase $\Theta_{K\p}$  already present
in the intermediate $K \p$ amplitude. 
Therefore $\Theta_L$ represents the gap between the two and three-body phases, which depends on both $m_{12}$ and $m_{23}$, showing that Watson's theorem does not apply to this case.
Our results both confirm  the dominance of weak vector currents in 
this branch of $D^+$ decays and indicate that
proper three body final-state 
interactions, in which the initial four-momentum of the $D^+$ 
is shared among all final particles, are rather important
over the whole energy range considered.
In a parallel study, to be presented elsewhere,
we found that this feature is also present in 
the $P-$wave projection of final-state $K \p$ subamplitude,
which has  a non-vanishing phase at threshold.  

\section*{ACKNOWLEDGEMENTS}
The authors  thanks A. dos Reis, I. Bediada and T. Frederico for fruitful discussions. This work was supported by Funda\c{c}\~{a}o de Amparo \`{a} Pesquisa do Estado de S\~{a}o Paulo (FAPESP). 

\appendix
\section{kinematics}
\label{kinematics}

The momentum of the $D$-meson is $P\,$, whereas those of the outgoing
kaon and pions are $p_2\,$, $p_1\,$, and $p_3\,$, respectively. 
The invariant masses read
\bea
&& \hspace*{-15mm}
m_{12}^2= (p_1 \sp p_2)^2 = M_\p^2 + M_K^2 + 2\,p_1 \cd p_2 \;,
\label{A.1}\\[2mm]
&& \hspace*{-15mm}
m_{13}^2= (p_1 \sp p_3)^2 = 2\,M_\p^2 + 2\,p_1 \cd p_3 \;,
\label{A.2}\\[2mm]
&& \hspace*{-15mm}
m_{23}^2= (p_2 \sp p_3)^2 = M_\p^2 + M_K^2 + 2\,p_2 \cd p_3 \;,
\label{A.3}
\eea
and satisfy the constraint
\beq
M_D^2 = m_{12}^2 + m_{13}^2 + m_{23}^2 - 2\,M_\p^2 - M_K^2 \;.
\label{A.4}
\eeq
The projection into partial waves for subsystem $(12)$
is performed by going to its center of mass and writing
\bea
m_{13}^2 &\!=\!& \a_{13}^2 - \b_{12} \,\cos\theta \;,
\label{A.5}\\[2mm]
m_{23}^2 &\!=\!& \a_{23}^2 + \b_{12} \,\cos\theta \;,
\label{A.6}\\[2mm]
\a_{13}^2 &\!=\!& \lb M_D^2 + 2\,M_\p^2 + M_K^2 - m_{12}^2
- \, (M_D^2 \sm M_\p^2)\,(M_K^2 \sm M_\p^2)/m_{12}^2 \rb/2 \;,
\label{A.7}\\[2mm]
\a_{23}^2 &\!=\!& \lb M_D^2 + 2\,M_\p^2 + M_K^2 - m_{12}^2
+ \, (M_D^2 \sm M_\p^2)\,(M_K^2 \sm M_\p^2)/m_{12}^2 \rb/2 \;,
\label{A.8}\\[2mm]
\b_{12} &\!=\!& 2\, Q' q' \;,
\label{A.9}\\[2mm]
q' &\!=\!& \frac{1}{2\sqrt{m_{12}^2}}
\lb m_{12}^4 - 2(M_\p^2 \sp M_K^2)\,m_{12}^2 
+ (M_\p^2 \sm M_K^2)^2 \rb^{1/2}
\;,
\label{A.10}\\[2mm]
Q' &\!=\!& \frac{1}{2\sqrt{m_{12}^2}}
\lb m_{12}^4 - 2(M_\p^2 \sp M_D^2)\,m_{12}^2 
+ (M_\p^2 \sm M_D^2)^2 \rb^{1/2} \;,
\label{A.11}
\eea
where $\theta$ is the angle between the momenta of the pions.

\section{basic $D^+ \rar \Kb^0 \p^0 \p^+$ amplitude}
\label{basic}
Our description of the decay $D^+ \to K^-\pi^+\pi^+$
includes both the primary weak vertex
and hadronic final state interactions, associated with successive
$K\p$ scatterings.
When the $W\rar \p\p$ vertex is corrected by means of time-like
form factors, both the $\rho$-meson and $P$-wave $\p\p$ interactions 
also become part of the problem.
This could, in principle, give rise to a structure of 
final interactions depending  on both $\p\p$ and $K\p$
amplitudes.
For the sake of keeping complications under control,
we consider here just $\p\p$ interactions which occur to the 
left of $K\p$ amplitudes.
Therefore, the amplitude for the process 
$D^+(P) \rar \Kb^0(p_K) \p^0(p_0) \p^+(p_+)$,
given in Fig.\ref{Fro} and denoted by $A_0$, becomes 
the basic building block in the evaluation of the weak vector series.

We begin by constructing $T_{\p\p}^{P1}$, 
the isospin $I=1$, $P$-wave $\p\p$ amplitude.
The momenta of the outgoing pions are $p_+$ and $p_0\,$,
whereas those inside the two-pion loop are $q_+$ and $q_0$.
The total momentum is $Q=(p_+ + p_0)=(q_+ + q_0)$ and
the loop integration variable is $\ell=(q_+ - q_-)/2$.
Assuming that, at low energies, $\p\p$ interactions are dominated 
by a $\cO(q^2)$ contact term supplemented by the $\cO(q^4)$ 
$\rho$-pole contribution, the effective lagrangians
in Ref.\cite{EGPR}  yield the tree contribution
\bea
&& \Tb^1 = (t-u) 
\lb \frac{1}{F^2} - \frac{2\,G_V^2}{F^4}\, \frac{s}{s-m_\rho^2}\rb \,,
\label{B.1}
\eea 
where $F$ is the pion decay constant and $G_V$ describes the 
$\rho\,\p\p$ coupling.
The approximation $G_V = F/\sqrt{2} \sim 66\,$MeV yields 
a more compact structure, given by 
\bea
&& \Tb^1 = -\,\frac{(t-u)}{F^2} \; \frac{m_\rho^2}{s-m_\rho^2} \,.
\label{B.2}
\eea 
For free particles in the center of mass frame, 
$(t-u)=(s - 4M_\p^2)\, \cos \theta$ and $P$-wave
projection yields the kernel
\bea
\cK^{P1} = -\, \frac{(s-4 M_\p^2)}{3\, F^2} \; 
\frac{m_\rho^2}{s - m_\rho^2}
\label{B.3}
\eea

The iteration of this kernel by means of intermediate two-pion states
produces the unitarized amplitude\cite{OO}
\bea
T_{\p\p}^{P1} = \frac{\cK^{P1}}{1 + \cK^{P1} \,\Omega_{\p\p}} \;,
\label{B.4}
\eea
where $\Omega_{\p\p}$ is a divergent loop function.
Therefore, we write it as the sum of an 
infinite constant $\Lambda_\infty$
and a regular component $\Ob_{\p\p}\,$, given by\cite{BR}
\bea
&& \Ob_{\p\p} = - \frac{S}{16\p^2}\,
\lc 2 - \frac{\sqrt{\l}}{s}\; 
\ln\lb \frac{s- 2M_\p^2 + \sqrt{\l}}{2M_\p^2}\rb 
+ i\; \p \;\frac{\sqrt{\l}}{s} \rc \;,
\nn\\[4mm]
&& \l = s^2 - 4\,s\,M_\p^2 \;,
\label{B.5}
\eea
where $S=1/2$ is the symmetry factor for identical particles.
After regularization, one finds
\bea
T_{\p\p}^{P1} = \frac{\cK^{P1}}{1 + \cK^{P1}\,(\Ob_{\p\p} + C_{\p\p)}} \;,
\label{B.6}
\eea
where $C_{\p\p}$ is an arbitrary constant.
This amplitude is related with phase shifts by
\bea
T_{\p\p}^{P1} &\!=\!& 32 \p \, \lb \frac{s}{s-4M_\p^2} \rb^{1/2}
\sin\d \, e^{i\d}
\label{B.7}
\eea
and we fix $C_{\p\p}$  by the phase at $90^0\,$.
The I=1 amplitude to be used in the evaluation of $A_0$ is 
given by eq.(\ref{B.6}) multiplied by $(3\,\cos\theta)$.
It is denoted by $T_{\p\p}^1$ and can be cast in the covariant form
\bea
T_{\p\p}^1 = 3 \frac{(t-u)}{s - 4M_\p^2}\;T_{\p\p}^{P1}
= - \, 6 \,\frac{(p_+ - p_0)^\n \ell_\n}{s - 4 M_\p^2} \;T_{\p\p}^{P1} \;.
\label{B.8}
\eea

Going back to the decay amplitude $A_0$ and 
reading the diagrams in Fig.\ref{Fro}, one finds 
\bea
A_0 &\!=\!& [G_F\, \cos^2\theta_C]\, 
\la \Kb^0 |V^\m|D^+ \ra \,
\lb \frac{m_\rho^2}{Q^2 \sm m_\rho^2}\rb 
\lb \, (p_+ \sm p_0)_\m 
+ i\, 6 \, \frac{(p_+ - p_0)^\n }{Q^2 - 4 M_\p^2} \;T_{\p\p}^{P1}(Q^2) \;
I_{\m\n} \rb \,,
\nn\\[2mm]
I_{\m\n} &\!=\!& 
\int \! \frac{d^4 \ell}{(2\p)^4} \; 
\frac{\ell_\m\, \ell_\n}
{[(\ell \sp Q/2)^2 - M_\p^2]\,[(\ell \sm Q/2)^2 -M_\p^2]} \,,
\label{B.9}
\eea
where $G_F$ is the Fermi constant, $\th_C$ is the Cabibbo angle,
$V^\m$ is the weak vector current.
The regular part of $I_{\m\n}$ can be related with 
eq.(\ref{B.5}) and one has\cite{HR,PCM}
\bea
I_{\m\n} = \frac{i}{6} \lb Q^2 - 4 M_\p^2 \rb
\lb g_{\m\n} -\frac{Q_\m Q_\n}{Q^2} \rb \; [\Ob_{\p\p} + C_{\p\p}]
\label{B.10}
\eea
Using this result into eq.(\ref{B.9}) and recalling
that $(p_+ - p_0)^\n Q_\n =0$ for on shell particles,
one has 
\bea
A_0 &\!=\!& [G_F\, \cos^2\theta_C]\,
\la \Kb^0|V^\m |D^+\ra  \; (p_+ \sm p_0)_\m\;
\frac{m_\rho^2}{D_\rho} \;,
\label{B.11}\\[2mm]
D_\rho &\!=\!& 
(Q^2 \sm m_\rho^2) - (m_\rho^2/3F^2)\,(Q^2 \sm 4M_\p^2)\,[
\, [\Ob_{\p\p} + C_{\p\p}] \;.
\label{B.12}
\eea
The vector current matrix element is written as
\bea
\hspace*{-10mm}
\la \, \Kb^0 \,| \, V^\m \, | \, D^+ \, \ra 
&\!=\!& \lp P_D^\m + p_K^\m \rp F_1^{DK}(Q^2)
- Q^\m \; \frac{M_D^2 \sm M_K^2}{Q^2}
\lb F_1^{DK}(Q^2) - F_0^{DK}(Q^2) \rb 
\label{B.13}
\eea 
and form factors are parametrized in terms of 
vector and scalar $c\sb$ nearest poles as\cite{weakFF}
\bea
F_1(Q^2) = \frac{F^{DK}(0)}{1-Q^2/m_V^2}
\hspace{5mm} \mathrm{and} \hspace{5mm}
F_0(Q^2) = \frac{F^{DK}(0)}{1-Q^2/m_S^2} \,,
\label{B.14}
\eea
with  $V \rar D_s^*(2100)$, $S \rar D_s^*(2317)$
and $F^{DK}(0)=0.75$.

The denominator $D_\rho$ describes the $\rho$ meson and 
includes its dynamically generated width.
The function $D_\rho$ does not vanish along the
real axis, in spite of the bare $\rho$ propagators
in Fig.\ref{Fro}. 
It has a zero in the second Riemann sheet, 
quite close to the value quoted in Ref.\cite{CGL}, 
namely at $Q^2=(m_\rho - i\, \Gamma_\rho/2)^2\,$,
$m_\rho=762.4 \pm 1.8\,$MeV, $\Gamma_\rho=145.2 \pm 2.8\,$MeV. 

In order to simplify calculations one notes that the ratio 
$m_\rho^2/D_\rho$ in eq.(\ref{B.11}) is related to the 
$P$-wave amplitude given by eq.(\ref{B.4}) by
\bea
\frac{m_\rho^2}{D_\rho} = -\; \frac{3\,F^2}{s-4M_\p^2} \; 
T_{\p\p}^{P1} \;.
\label{B.15}
\eea
Using the data from Hyams et al.\cite{Hyams}, we fitted this 
ratio using the structure 
\bea
\frac{m_\rho^2}{D_\rho} = 
\frac{N_\rho}{s-\th_\rho} 
+ \frac{N_1}{s-\th_1}
+ \frac{N_2}{s-\th_2} \;,
\label{B.16}
\eea
with free parameters $\th_k=\th_{kR} + i\, \th_{kI}$ and 
$N_k=N_{kR} + i\, N_{kI}\,$, given in table \ref{BT}.

\begin{table}[h]
\begin{tabular} {|c|c|c|c|c|}
\hline
 $\;\;k\;\;$  & $\theta_R$ & $\theta_I$ & $N_R$ & $N_I$ \\ \hline 
$\rho$   & 0.580133 & -0.1137172 & 0.6131598 & -0.1107509 \\ \hline
1 & 2.539625 & -0.6468928 & 0.2418401 & -0.1080483 \\ \hline
2 & 3.642091 & -0.1595399 & 0.0016668 & -0.1941643 \\ \hline 
\end{tabular}
\caption{Fitted parameters in eq.(\ref{B.16})}.
\label{BT}
\end{table}
%

\begin{figure}[h] 
\begin{center} 
\includegraphics[width=0.45 \columnwidth,angle=0]{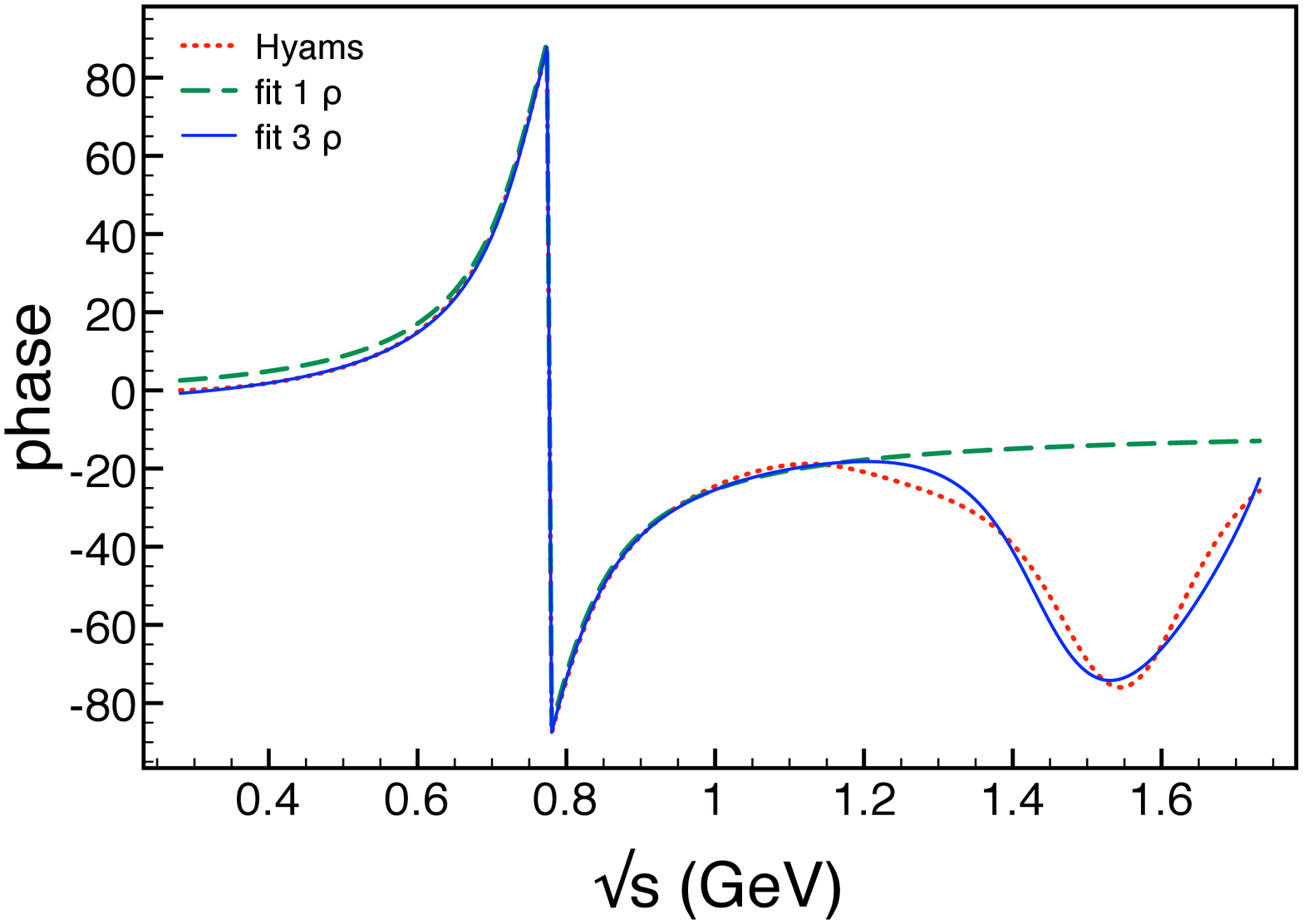}
\includegraphics[width=0.45 \columnwidth,angle=0]{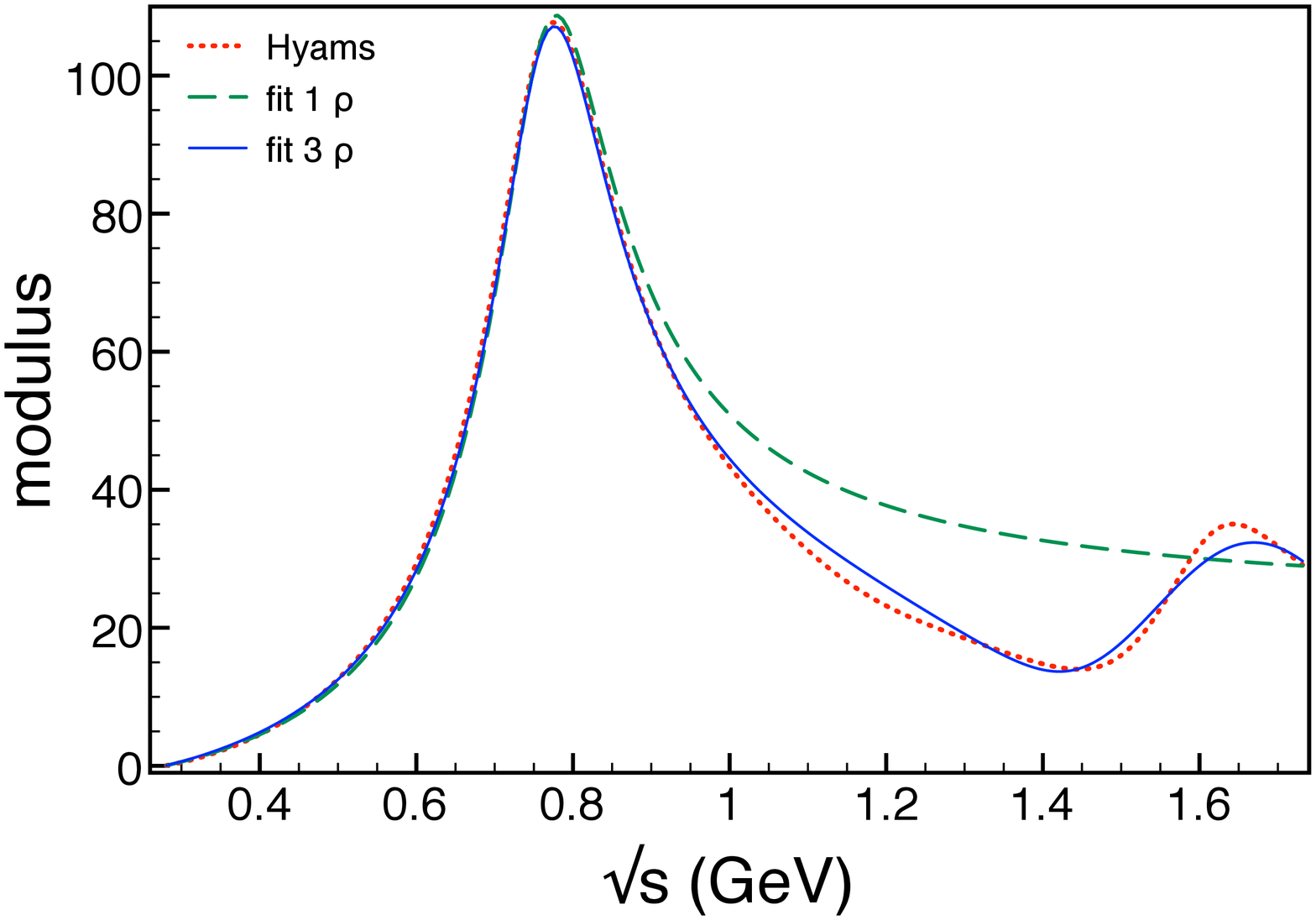}
\end{center}
\caption{
Results for $\p\p$ phase and modulo with only one-$\rho$ (dashed) and adding another two poles (continuous), compared with Hyams et al.\cite{Hyams} (dotted).}
\label{F11-a}
 \end{figure}

\begin{figure}[h] 
\begin{center}
\includegraphics[width=0.6 \columnwidth,angle=0]{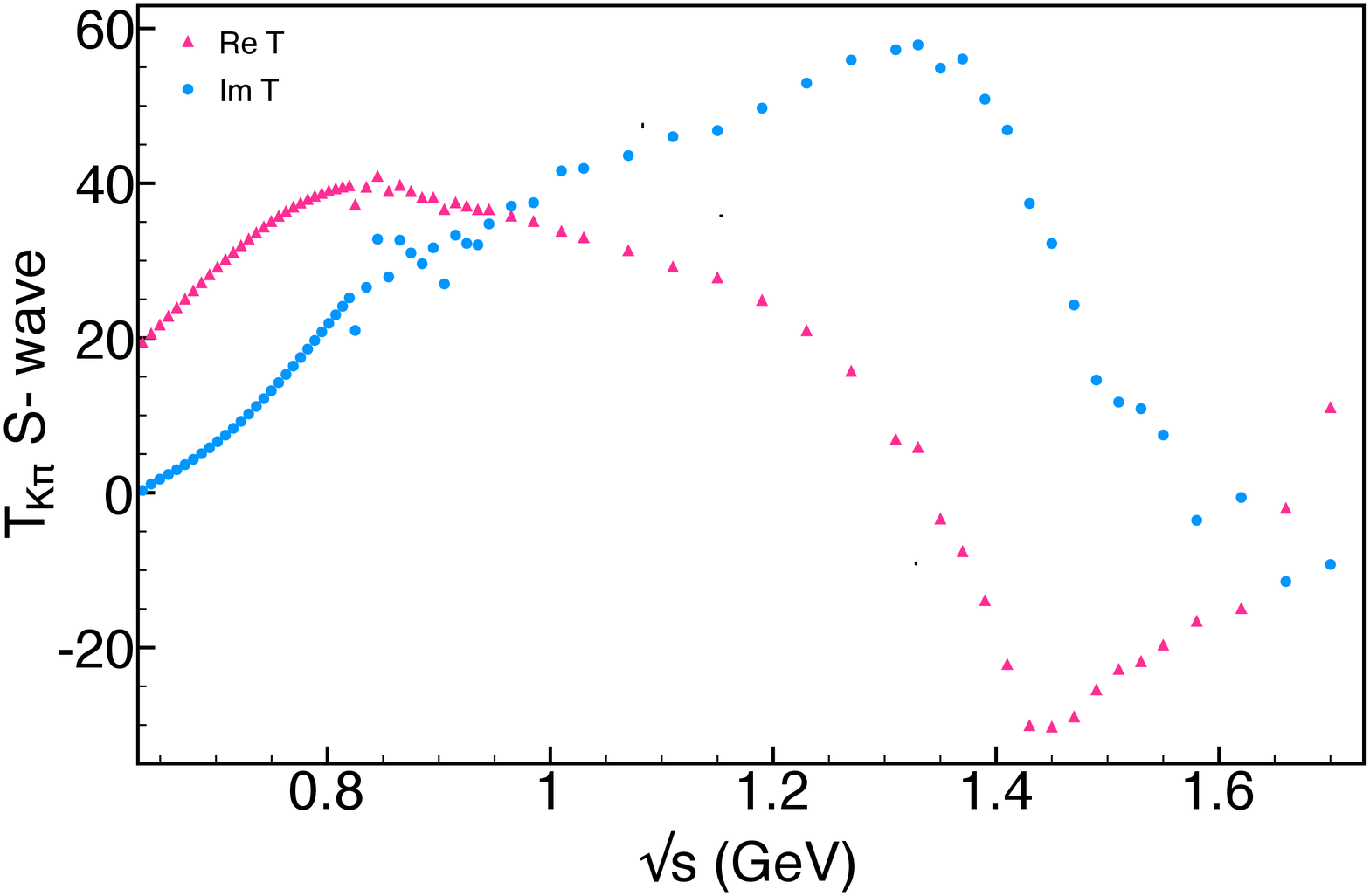}
\end{center}
\caption{
Function $m_\rho^2/D_\rho$; 
the red continuous curve represents eq.(\ref{B.15}) 
with parameters from Hyams et al.\cite{Hyams}
and the black dotted curve is our fit, using eq.(\ref{B.16});
as data begin at $0.6\,$GeV, the red curve to the left of the 
vertical dashed line corresponds to an extrapolation.}
\label{F11-b}
\end{figure}

In Figs.~\ref{F11-a} and \ref{F11-b} we display  the importance of the inclusion of higher poles in eq.(\ref{B.16}) in extending the agreement  with Hyams et. al.\cite{Hyams} data.

The expression for $A_0$ to be used in calculations is obtained
by assembling previous results, and one finds
\bea
A_0 &\!=\!& -\,[G_F\, \cos^2\theta_C \,F^{DK}(0)]\,
\lb \frac{N_\rho}{Q^2-\th_\rho} 
+ \frac{N_1}{Q^2-\th_1}
+ \frac{N_2}{Q^2-\th_2} \rb
\nn\\[2mm]
&\! \times \!&
\lc \lb M_D^2 \sp 2M_\p^2 \sp M_K^2 - 2(p_0 \sp p_K)^2
-Q^2 + (p_0^2 \sm M_\p^2) + (p_K^2 \sm M_K^2) \rb \, 
\frac{m_V^2}{Q^2 \sm m_V^2} \right.
\nn\\[2mm]
&\!-\!& \left.
(p_0^2 \sm M_\p^2)\,
\frac{(M_D^2 \sm M_K^2)\,(m_V^2 \sm m_S^2)}
{(Q^2 \sm m_V^2)\,(Q^2 \sm m_S^2)} \rc \;.
\label{B.17}
\eea

\newpage
\section{$K\p$ amplitude}
\label{kapi}

In this work, one needs the elastic $K\p$ amplitude over 
the full Dalitz plot.
As there are no data\cite{LASS} available in the interval
$0.401\,$GeV$^2 \leq s \leq 0.681\,$GeV$^2$,
one encompasses this region with the help 
of a theoretical amplitude, based on the unitarized chiral symmetry.
This model has been discussed in detail in Ref.\cite{BR,PAT-Kpi}
and here we just summarize its main features.

For each spin-isospin channel, the unitary amplitude $T_{LI}$ 
is obtained by ressumming the infinite geometric series 
\bea
T_{LI} &\!=\!& \cK_{LI} - \cK_{LI}\,[\Ob_{K\p} \sp C_{LI}] \,\cK_{LI}
+ \cK_{LI}\,[\Ob_{K\p} \sp C_{LI}] \, \cK_{LI}\,[
\Ob_{K\p} \sp C_{LI}] \,\cK_{LI} 
+ \cdots 
\nn\\[4mm]
&\!=\!& \frac{\cK_{LI}}{1 + [\Ob_{K\p} \sp C_{LI}] \,\cK_{LI}} \;,
\label{C1}
\eea
where $\cK_{LI}$ is a kernel
and the function $\Ob_{K\p}$, related with 
the two-meson propagator, is given by\cite{BR} 
\bea
\Ob_{K\p} &\!=\!& 1+ 
\frac{M_\p^2 \sp M_K^2}{M_\p^2 \sm M_K^2} \, \ln \frac{M_\p}{M_K}
- \frac{M_\p^2 \sm M_K^2}{s} \, \ln \frac{M_\p}{M_K}
\nn\\[2mm] 
&\!-\!& \frac{\sqrt{\l}}{s}\,
\ln \lb \frac{s \sm M_\p^2 \sm M_K^2 + \sqrt{\l}}{2 M_\p M_K} \rb 
+ i \, \p\, \frac{\sqrt{\l}}{s} \;,
\nn\\[2mm]
\l &\!=\!& s^2 -2s(M_\p^2 \sp M_K^2) + (M_\p^2 \sm M_K^2)^2 \;,
\label{C2}
\eea
and $C_{LI}$ is a constant.
Chiral perturbation theory determines the kernels $\cK_{LI}$
as the sum of a $\cO(q^2)$ contact 
term\cite{GL}, supplemented by $\cO(q^4)$ corrections,
which we assume to be dominated by $s$-, $t$- and $u$-channel 
resonances\cite{EGPR}.
In order to fit LASS data\cite{LASS}, we also included a higher 
mass resonance, as described in Ref.\cite{PAT-Kpi}.

\begin{figure}[h]
\hspace*{-25mm}
\includegraphics[width=0.35\columnwidth,angle=0]{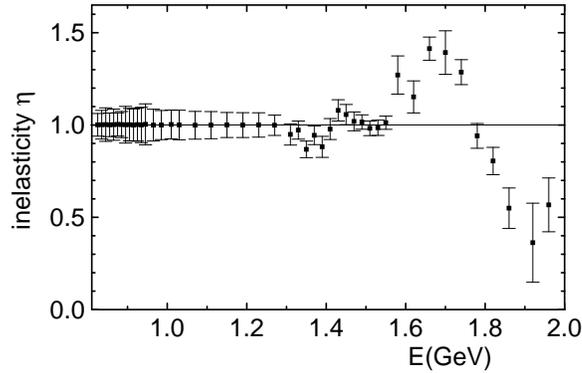}
\caption{Inelasticity parameter $\eta$ for $S_{1/2}$ LASS data.}
\label{FC1}
\end{figure}

In the case of the $S_{1/2}$ wave ($L,I=0,1/2$), the theoretical 
kernel is written as 
$\cK_{S_{1/2}} = \cK_{BG} + \cK_H \,$,
where $\cK_{BG}$ is a real background and $\cK_H$ includes resonances.
The former is given by $\cK_{{BG}} = \cK_C + \cK_S + c_V\,\cK_V\,$, with
\bea
&& \cK_C =
\frac{1}{F^2} \,\lb s - 3\, \bq^2/2 - (M_\p^2 \sp M_K^2)\rb \;,
\label{C3}\\[4mm]
&& \cK_S = -\, \frac{4}{F^4}
\lc \lb \ct_d^2 \,m_0^2 - 2  \ct_d \, (\ct_d \sm \ct_m)(M_\p^2 \sp M_K^2) 
- 2 \ct_d^2 \,\bq^2 \rb \right.
\nn\\[2mm]
&& \left.
+ \lb \ct_d \, m_0^2 - 2(\ct_d \sm \ct_m)M_\p^2 \rb
\lb \ct_d m_0^2 - 2(\ct_d \sm \ct_m)M_K^2 \rb \,I_S^t(\bq^2; m_0^2) \rc
\nn\\[2mm]
&& + \; \frac{1}{3F^4}
\lc \lb c_d^2 \, m_8^2 - 2 c_d(c_d \sm c_m)(M_\p^2 \sp M_K^2) 
- 2 c_d^2\, \bq^2 \rb \right.
\nn\\[2mm]
&& \left.
+ \lb c_d \, m_8^2 - 2(c_d \sm c_m)M_\p^2 \rb
\lb c_d \, m_8^2 - 2(c_d \sm c_m)M_K^2 \rb I_S^t(\bq^2 ; m_8^2) \rc
\nn\\[4mm]
&& + \frac{1}{2F^4}\;
\lc \lb c_d^2\,(m_{K_0^*}^2 \sp 2 M_\p^2 \sp M_K^2 \sm s \sp 2\,\bq^2)
+ 2 c_d (c_d \sm c_m) \,(M_\p^2 \sp M_K^2)\,\rb \right.
\nn\\[2mm]
&& \left. + \lb c_d \,m_{K_0^*}^2 
\sm  (c_d \sm c_m) \,(M_\p^2 \sp M_K^2) \rb^2 \, 
I_S^u(\bq^2;m_{K_0^*}^2 ) \rc \;,
\label{C4}\\[4mm]
&& \cK_V = - \lb \frac{G_V}{F^2} \rb^2
\lc \lb 2(s \sm M_\p^2 \sm M_K^2) + m_\rho^2 -2\bq^2 \rb \right.
\nn\\[2mm]
&& \left. 
+ m_\rho^2 \, \lb m_\rho^2 \sp 2 \,(s \sm M_\p^2 \sm M_K^2) \rb \,
I_S^t(\bq^2;m_\rho^2) \rc
\nn\\[4mm]
&& - \frac{1}{4} \, \lb \frac{G_V}{F^2} \rb^2
\lc \lb m_{K^*}^2 \sp s  \sp 2 \bq^2\rb \right.
\nn\\[2mm]
&& \left. 
+ \lb m_{K^*}^2 \lp m_{K^*}^2 \sp 2 \, (s \sm M_\p^2 \sm M_K^2) \rp
\sm (M_\p^2 \sm M_K^2)^2 \rb \, I_S^u(\bq^2; m_{K^*}^2) \rc \;,
\label{C5}\\[4mm]
&& I_S^t(\bq^2; m^2) = -\,\frac{1}{4\bq^2}\, 
\ln \lb 1 + \frac{4\bq^2}{m^2}\rb \;,
\label{C6}\\[4mm]
&& I_S^u(\bq^2; m^2) = \frac{1}{4\bq^2}\,
\ln \lb 1 - \frac{4\bq^2}{m^2 \sp s\sm 2(M_\p^2 \sp M_K^2)} \rb \;,
\label{C7}
\eea
where $F$, $c_d, c_m, \cb_d, \cb_m$, and $G_V$ are coupling constants
and the CM three-momentum is 
\bea
\bq^2 &\! = \!& \frac{1}{4 \,s} \;
\lb s^2 - 2s\,(M_\p^2 \sp M_K^2) + (M_\p^2 \sm M_K^2)^2 \rb \;.
\label{C8}
\eea
Two $s$-channel resonances are incorporated as sum of 
Breit-Wigner functions\cite{PAT-Kpi} 
\bea
\cK_H &\!=\!& -\,\frac{3}{2F^4} 
\lb \frac{\lb c_d \, s - (c_d \sm c_m) \, (M_\p^2 \sp M_K^2) \rb^2}
{s- m_{K_0^*}^2 + i\,g_a^2\, Q_a/8\p\sqrt{s} } 
+ \frac{\lb c_{db} \, s - (c_{db} \sm c_{mb}) \, (M_\p^2 \sp M_K^2) \rb^2}
{s-m_b^2 + i\,g_b^2\,Q_b/8\p\sqrt{s}} \rb \;,
\label{C10}\\[2mm]
g_i &\!=\!& A_i + B_i \, s \;,
\label{C11}
\\[2mm]
Q_i &\!=\!& \frac{\sqrt{s}}{2}\; (1-h_i^2/s) \;.
\label{C12}
\eea
The usual inelasticity parameter $\eta\,$, evaluated for $S_{1/2}$ data, 
is shown in Fig.\ref{FC1}.
Points for which $\eta>1$ within error bars were discarded in our fit.

We have extended $S_{1/2} \, K\p$ data to threshold by means of two different fits. 
The first one includes just a single resonance and holds for energies smaller than
$1.33$GeV, whereas the second one includes two resonances and is valid over the whole Dalitz plot.
Their correspond respectively to $\chi^2/n.d.f.\,=\, 0.55$ and  $\chi ^2/n.d.f.\,=\, 1.62$. 
Our parameters, in suitable powers of GeV, are: $F \!=\! 1.02722, \,G_V  \!=\! 0.0686287$,$m_8  \!=\! m_0  \!=\!  0.983$ and $C_{S_{1/2}}  \!=\! 1.124899\times 10^{-2}$,
$m_{K_0^*} \!=\! 1.108858 $, 
$c_d \!=\! 0.0254505, \, c_m \!=\! 0.1483455$, 
$A_a \!=\! 4.563646, \, B_a \!=\! -2.055842, \, h_a \!=\! 1.138489, \, c_V=0.26200$ 
for the single resonance fit and 
$C_{S_{1/2}}  \!=\!  -2.273182\times 10^{-3}$,
$m_{K_0^*} \!=\! 1.338404 $, 
$c_d \!=\! 0.026607, \, c_m \!=\! 0.017428$, 
$A_a \!=\! 4.952313, \, B_a \!=\! -1.956429, \, h_a \!=\! 1.130126$,
$m_b \!=\! 2.003338, \, c_{db} \!=\! 0, \, c_{mb} \!=\! 0.166268$, 
$A_b \!=\! 5.042537, \, B_b \!=\! -7.182061, \, h_b\!=\!1.809129$,
and $c_V=0.89272$, for the two-resonance case.

\begin{figure}[h]
\begin{center}
\hspace*{-10mm}
\includegraphics[width=0.46\columnwidth,angle=0]{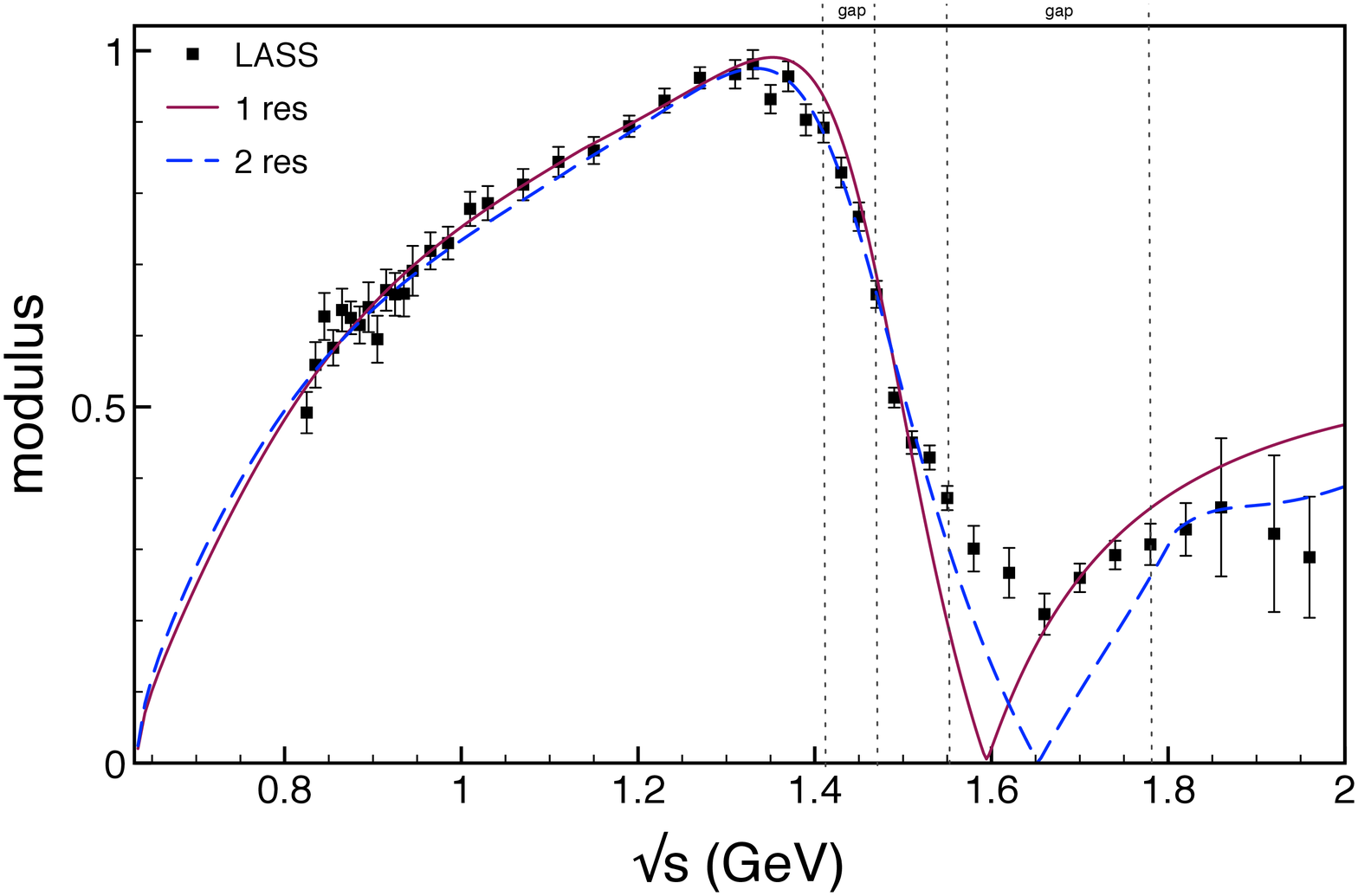}
\hspace*{10mm}
\includegraphics[width=0.46\columnwidth,angle=0]{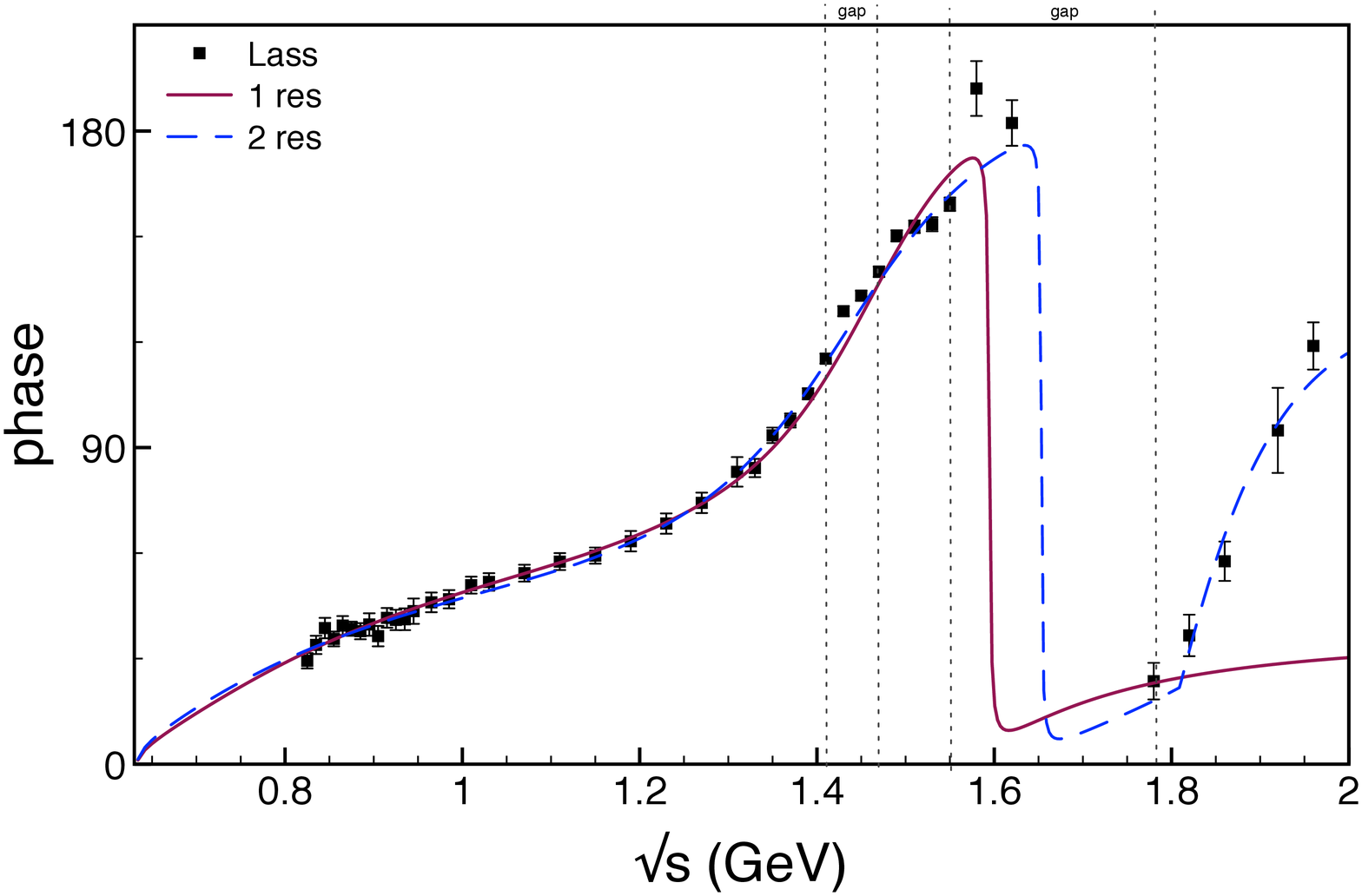}
\end{center}
\caption{Fits for the modulus and phase of the $K\p$ $S_{1/2}$ LASS data;
points within the regions indicated as {\em gap} in the top axis 
were excluded from the fit.}
\label{FC2}
\end{figure}

Both fits for the modulus an phase are given in Fig.\ref{FC2}. In the \dkpp decay amplitude, alternatively, we can use directly empirical data from LASS\cite{LASS}  and  merge it with the low energy fit, where there is no data. This became what we called {\it hybrid} amplitude.  

\begin{figure}[h]
\includegraphics[width=0.6\columnwidth,angle=0]{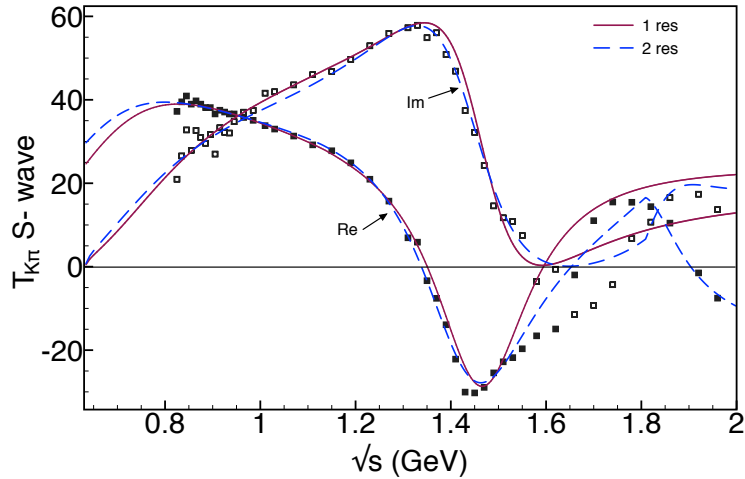}
\caption{Real and imaginary components of the $S_{1/2}$ $K\p$ amplitude
fitted to LASS data (squares) and extended to low-energies using chiral 
symmetry.}
\label{FC3}
\end{figure}

%
In Fig.\ref{FC3} we show the real and imaginary components of the amplitude. 
One notice that values for the real part at threshold are different, 
namely 24 and 30, and they can be compared with those  obtained by ChPT\cite{Bernard} 
and dispersion relations\cite{Moussalam}, respectively T = 21.7 and T = 25.5. 
These values indicate that the low energy fit is more suitable to describe low energy behaviour. 
%
It is worth noting that both real and imaginary components are very small for 
 $\sqrt{s}\sim 1.65\,$GeV in the two-resonance result.

\section{loop integrals}
\label{int}

\bea
I_{abc}^S &\!=\!& \int \frac{d^4 \ell}{(2\p)^4} \;
\frac{16 \p^2 }{D_a \, D_b \, D_c } \;,
\;\;\;
I_{abcd}^S = \int \frac{d^4 \ell}{(2\p)^4} \;
\frac{16 \p^2 }{D_a \, D_b \, D_c \, D_d } \;.
\nn
\eea

We begin by discussing the integrals $I^S$, given by eqs.(\ref{2.7}).
Their treatment can be simplified because the $\rho$ and 
the $c\sb$ state entering the form factor share the 
same momentum. 
This allows, for instance, one to write
\bea
I_{\p K \rho V}^S &\!=\!& \frac{1}{M_V^2 \sm \T_R + i \T_I}
\lb I_{\p K V}^S - I_{\p K \rho}^S \rb \;,
\label{D1}
\eea
where $\T$ is the parameter defined in App.\ref{basic}. 
Similar simplifications can be performed every time subscripts
$\rho V$ or $\rho S$ occur.

The integral $I_{\p K \rho}^S$ is important in this
problem because its imaginary part is determined by two different
thresholds, associated with cuts along $K \p$ and $K \rho$
propagators.
Using results from App.\ref{basic}, one writes
\bea
I_{\p K \rho}^S &\!=\!& \int \frac{d^4 \ell}{(2\p)^4} \;
\frac{16 \p^2}{[(\ell \sm p_3)^2 \sm M_\p^2] \, 
[(\ell \sm P)^2 \sm M_K^2]} \; 
\frac{N_\rho}{\ell^2 - \T_R + i \T_I} \;.
\label{D2}
\eea
Representing this function by means of Feynman parameters and performing
one of the integrals analytically, one finds
\bea
I_{\p K \rho}^S &\!=\!&  i\,N_\rho \; \P_{\p K \rho}\;,
\label{D3}\\[2mm]
\P_{\p K \rho} &\!=\!& - \int_0^1 \!\! d a\; J_{\p K \rho}(a) \;,
\label{D4}
\eea
with
\bea
J_{\p K \rho}(a) &\!=\!&  \frac{1}{\sqrt{\l}} 
\lc \lb \; \ln\frac{|F_1|\,|G_2|}{|G_1| \, |F_2|}\;\rb 
+ i\, \lb \; \theta_{J} \;\rb \rc \;.
\label{D5}\\[2mm]
\theta_{J} &\!=\!& 
\lb \; \theta_{F1} - \theta_{F2} - \theta_{G1} + \theta_{G2} \; \rb  \;,
\label{D6}\\[2mm]
F_{1,2} &\!=\!& 
\frac {|2M_D^2 \,a + B \mp \sqrt{\l}|}{M_D^2} \; e^{i\,\theta_{F1,2}} \;,
\label{D7}\\[2mm]
G_{1,2} &\!=\!& 
\frac{| B \mp \sqrt{\l}|}{M_D^2} \; e^{i\,\theta_{G1,2}} \;,
\label{D8}
\eea
and
\bea
B &\! = \!& \lb \T_R - i\T_I - M_\p^2 - M_K^2 + m_{12}^2 
- a\,(M_D^2 - M_\p^2 + m_{12}^2) \rb   \;,
\label{D9}\\[2mm]
\l &\!=\!& B^2 - 4\, M_D^2\, C \;,
\label{D10}\\[2mm]
C &\! = \!& \lb (1 \sm a)\,M_\p^2 + a\, M_K^2 
- a\, (1 \sm a)\,m_{12}^2 \, \rb  \;.
\label{D11}
\eea
The $\rho$ width is incorporated into the factors $N_\rho$, $\T_I\,$, 
and the case of a point-like resonance is recovered by making 
$N_\rho \rar 1\,$, $\T_I \rar \e\,$.

The integral $I_{\p K V}^S$ is 
\bea
I_{\p K V}^S &\!=\!& \int \frac{d^4 \ell}{(2\p)^4} \;
\frac{16 \p^2}{[(\ell \sm p_3)^2 \sm M_\p^2] \, 
[(\ell \sm P)^2 \sm M_K^2]} \; 
\frac{1}{\ell^2 - m_V^2} 
= i\,  \P_{\p K V}\;,
\label{D12}
\eea
and its evaluation is totally similar.
However, as now $m_V>M_D\,$, its imaginary part comes just from the
cut of the diagram along the $K\p$ subsystem. 
Integrals $I_{\p \rho V}^S$, $I_{K \rho V}^S$ 
and $I_{K \rho S}^S$ do not depend on $m_{12}^2$.


\end{document}